\documentclass{optica-article}

\journal{opticajournal} 

\articletype{Research Article}

\usepackage{lineno}

\begin{document}
	
	\title{An experimental technique for measuring radial coherence}
	
	\author{Radhika Prasad,\authormark{1,*} Nilakshi Senapati,\authormark{1} Suman Karan,\authormark{1} Abhinandan Bhattacharjee,\authormark{1,2} Bruno Piccirillo,\authormark{3} Miguel A. Alonso,\authormark{4,5} and Anand K. Jha\authormark{1,*}}
	
	\address{\authormark{1}Department of Physics, Indian Institute of Technology Kanpur, Kanpur, UP 208016, India\\
		\authormark{2}Currently at Universität Paderborn, Warburger Strasse 100, 33098 Paderborn, Germany\\
		\authormark{3}Dipartimento di Fisica, Università di Napoli Federico II, Complesso Universitario di Monte Sant'Angelo, via Cintia, Napoli, 80126, Italy\\
		\authormark{4}Aix Marseille Univ, CNRS, Centrale Med, Institut Fresnel, F-13013 Marseille, France\\
		\authormark{5}The Institute of Optics, University of Rochester, Rochester, New York 14627, USA}
	
	\email{\authormark{*}radhikap@iitk.ac.in} 
	\email{\authormark{*}akjha@iitk.ac.in}
	
	
	\begin{abstract*} 
		Coherence refers to correlations between field vibrations at two separate points in degrees of freedom such as space, time, and polarisation.  In the context of space, coherence theory has been formulated between two transverse positions which can be described either in the cartesian coordinates or in the cylindrical coordinates. When expressed in cylindrical coordinates, spatial coherence is described in terms of azimuthal and radial coordinates. The description of spatial coherence in radial degree of freedom has been formulated only recently in JOSA A \textbf{40}, 411 (2023). In the present article, we demonstrate an efficient experimental technique for measuring radial coherence, and we report measurement of radial coherence of two different types of radially partially coherent optical fields. 
	\end{abstract*}
	

\section{Introduction}
	
Coherence theory is used for studying and characterising correlations between field vibrations at two separate space-time points in an electromagnetic field \cite{mandelwolf1995,goodman2015,bornwolf2013,wolf2007}. Optical fields with partial spatial coherence have found several applications including enhanced imaging in the presence of scattering and turbulent media \cite{redding2012natphot,redding2015pnas, bhattacharjee2020pra}, optical coherence tomography \cite{karamata2004optlett, kim2005journalofbiomedical, bhattacharjee2020optlett}, and optical communication \cite{pang2018optlett, bhattacharjee2020optlett}. Theory of partial coherence has been formulated for various degrees of freedom including polarisation \cite{wolf2007, patoary2019josab, meher2020josab, gori2006optlett}, space \cite{bornwolf2013, bhattacharjee2018apl,aarav2017pra, partanen2013aopt} and time \cite{kohl2001prl,leppanen2017photres}. In the context of space, coherence theory has been formulated between two transverse positions and is referred to as transverse spatial coherence. The transverse spatial position can be described either in Cartesian coordinates $(x, y)$ or in the cylindrical polar coordinates  $(r, \theta)$. Coherence theory in Cartesian coordinates  has been extensively studied in the last several decades \cite{mandelwolf1995,goodman2015,bornwolf2013, wolf2007, bhattacharjee2018apl}. When expressed in cylindrical coordinates, the spatial coherence can be expressed in terms of radial and azimuthal coordinates. In the azimuthal coordinate, a measure of coherence based on averaging over the radial coordinate has been particularly suitable for several experiments \cite{jha2011pra, kulkarni2017natcomm, kulkarni2018pra}. More recently, the theory of coherence has been formulated for radial variables
 \cite{bhattacharjee2023josaa, olvera2023jopt, santarsiero2018optlett, santarsiero2017optlett}. In particular, Ref.~\cite{bhattacharjee2023josaa} defines a measure of coherence based on averaging over the azimuthal variables.  A representation of generic cross-spectral densities in terms of the radial variable and the orbital angular momentum (OAM) has also been proposed recently \cite{korotkova2021pra}.

Efficient and accurate measurement of spatial coherence function can give rise to several applications. An important example is the well-known Hanbury Brown-Twiss (HBT) interferometer, which measures the modulus of the degree of spatial coherence of an optical field in a manner that is completely insensitive to phase fluctuations. Consequently, very accurate measurements of the angular diameter of stars have become possible even in the presence of atmospheric turbulence \cite{brown1956nature}. Therefore, developing experimental techniques for the efficient measurement of coherence is a very active field of research. For measuring spatial coherence, several efficient techniques have been demonstrated in recent times \cite{bhattacharjee2018apl, halder2020optlett, torcalMilla2023optlett}. Enhanced techniques for measuring angular coherence have resulted in very accurate measurement of the orbital angular momentum (OAM) spectrum \cite{kulkarni2017natcomm, kulkarni2018pra} as well as of certain class of quantum states in the OAM basis  \cite{kulkarni2020pra, karan2023prapp}. Just as the angular degree of coherence provides information on the OAM spectrum, the radial degree of coherence can be used to get the radial mode spectrum \cite{bhattacharjee2023josaa, choudhary2018optlett}. Combining both the radial and the angular coordinates in a Laguerre-Gaussian (LG) beam can open up new avenues for communication \cite{trichili2016scirep}. The radial index of the LG beam also shows better resilience to certain forms of turbulence \cite{wang2024optcomm}.

Although an experimental technique based on Mach-Zehnder interferometer was proposed in Ref.~\cite{bhattacharjee2023josaa} for measuring radial coherence, there has been no experimental demonstration of the technique. In the present article, we propose and demonstrate an experimental technique based on a common-path interferometer for measuring the radial coherence of partially coherent field. We report measurements of radial coherence of two different types of radially partially coherent optical fields. Ours is the first experimental demonstration of the measurement of radial coherence, and we expect it to be an enabler for applications based on radial coherence.

	\section{Theory}
	\subsection{Definition}

	Consider $E(x,y)$ as the electric field amplitude of a planar monochromatic light source at the transverse position $(x,y)$. This field can be expressed in the polar coordinate system as  $E(r,\theta)$, where $x\equiv r\cos\theta$ and $y\equiv r\sin\theta$. The corresponding cross-spectral density function $W(r_1,\theta_1,r_2,\theta_2)$ quantifies the field correlations between $(r_1,\theta_1)$ and $(r_2,\theta_2)$ and is defined by:
	\begin{equation}
		W\left(r_1,\theta_1,r_2,\theta_2\right) =\langle E^{*}(r_1,\theta_1)E(r_2,\theta_2) \rangle_e,
	\end{equation} 
	where $\langle ... \rangle_e$ represents the ensemble average. To quantify the correlations between radial positions $r_1$ and $r_2$, independent of their azimuthal coordinates, we take average over azimuthal coordinate. This yields the radial cross-spectral density function $W_R(r_1,r_2)$ \cite{bhattacharjee2023josaa}, defined as:
	\begin{equation}
		W_R\left(r_1, r_2 \right) \equiv \frac{1}{2 \pi} \int_{-\pi}^\pi W\left(r_1, \theta, r_2, \theta\right) \mathrm{d} \theta.
	\end{equation}
	Further, we define the complex degree of radial coherence between $r_1$ and $r_2$ as
	\begin{equation}\label{mu_eq}
		\mu_R\left(r_1, r_2\right)=\frac{W_R\left(r_1, r_2\right)}{\sqrt{I_R\left(r_1\right) I_R\left(r_2\right)}},
	\end{equation}
	where $I_R(r)=W_R(r, r)$ is the radial intensity distribution.
	The magnitude $|\mu_R(r_1,r_2)|$ quantifies the degree of coherence: $|\mu_R(r_1,r_2)|=1$ indicates the perfect coherence, $0<|\mu_R(r_1,r_2)|<1$ corresponds to partial radial coherence, and  $|\mu_R(r_1,r_2)|=0$ signifies no radial coherence.

	\subsection{Radially partially coherent fields: Incoherent mixture of Laguerre-Gaussian (LG) modes} \label{sec:lg}
	\begin{figure*}[t]
		\includegraphics[width =\columnwidth]{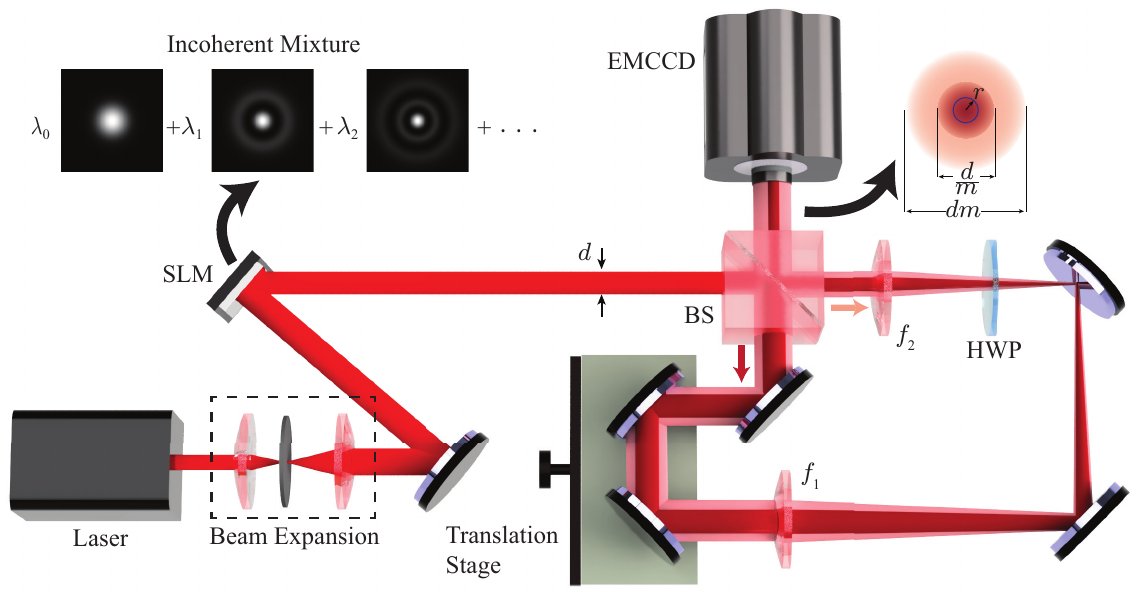}
		\caption{\label{fig:setup}Experimental setup involving common-path Sagnac interferometer. SLM: spatial light modulator; BS: beam splitter; HWP: half-wave plate; EMCCD: electron multiplying charge coupled device camera. The SLM produces a given radially partially coherent field (as an incoherent mixture of perfectly coherent radial modes), and the interferograms are recorded on the camera when HWP is kept with the optic axis orientations at $0$ and $\pi/4$ radians from the horizontal. We represent the reflected and transmitted beams at the BS with different colours (dark red and light red, respectively) to highlight the magnification due to $4f$ lens configuration clearly.}
	\end{figure*}
	
We first consider the class of partially coherent fields that have coherent mode representation in the Laguerre-Gaussian (LG) basis. The expression for LG mode $\mathrm{LG}_p^{l=0}(r)$ \cite{milonni2010,allen1992pra} with orbital angular momentum (OAM) index $l=0$ is given by
$
\mathrm{LG}_p^{l=0}(r)=A L_p^{|l|=0}\left[2 r^2 / w^2\right] \exp \left[-r^2 / w^2\right],
$
where $w$ is the beam waist, $p$ is the radial index, and $A$ is the normalisation constant. $L_p^{|l|=0}\left(2 r^2 / w^2\right)$ denotes the generalised Laguerre polynomial. Such a mode is radially perfectly coherent, having a degree of radial coherence of unity. If we consider an incoherent mixture of LG modes, having different proportion $\lambda_p$ corresponding to $p$ index, then we get a radially partially coherent field, the radial cross-spectral density function of which is given as:
	\begin{equation}
		W_R(r_1, r_2)=\sum_{p=0}^{p_{\max }} \lambda_p \mathrm{LG}_p^{* l=0}(r_1) \mathrm{LG}_p^{l=0}(r_2).\label{eqn:lg}
	\end{equation}

	\subsection{Radially partially coherent fields: Incoherent mixture of radial Hermite-Gaussian (HG) modes} \label{sec:hg}

We next consider the class of partially coherent fields that have coherent mode representation in the radial Hermite-Gaussian (HG) basis. If we consider an incoherent mixture of radial HG beams, then we get a radially partially coherent field. In analogy with the transverse Gaussian Schell Model (GSM) beams having partial transverse spatial coherence \cite{starikov1982josa,carter1977josa,he1988optcomm,deschamps1983josa}, we have a radial Gaussian Schell model (RGSM) field. The cross spectral density of the RGSM field is given by \cite{bhattacharjee2023josaa}: 
	$W_R\left(r_1, r_2\right)=\exp \left[-\frac{r_1^2+r_2^2}{4 \sigma_s^2}\right] \exp \left[-\frac{\left(r_2-r_1\right)^2}{2 \sigma_{\text {rad }}^2}\right]$
	where $\sigma_s$ is the radial beam size and $\sigma_{\mathrm{rad}}$ is the radial coherence width. The above form of the RGSM field can be realised by having an incoherent mixture of radial HG modes, given as:
\begin{equation}
	W(r_1,r_2)=\sum_{p=0}^{p_{\max}}\lambda_p HG^{*}_p(r_1)HG_p(r_2).\label{eqn:hg}
\end{equation}
Here, HG mode is given as $HG_{p}(r)=A H_p\left[\frac{\sqrt{2}r}{w}\right] \exp\left[-r^2/w^2\right]$. $H_p\left[\frac{\sqrt{2}r}{w}\right]$ denotes the Hermite polynomial. $A$ is the normalisation constant, $w$ is the beam waist, $p$ is the radial index, the coefficient $\lambda_p$ in the mixture is given by $\lambda_p=\tfrac{1}{\left[(Q/2)^2+1+Q\sqrt{(Q/2)^2+1}\right]^p}$, and $Q=\sigma_{\rm{rad}}/{2.25\sigma_s}$.	
One can control the $\sigma_s$ and $\sigma_{\mathrm{rad}}$ of the RGSM by modifying $Q$. The GSM field is a widely used model for partially coherent fields because of its Gaussian profiles for both the intensity distribution and degree of coherence function.
	
	\subsection{Proposed measurement scheme} \label{sec:scheme}

Figure \ref{fig:setup} illustrates the schematic of the proposed setup for measuring the radial cross-spectral density function of the incoming field. The scheme employs a 4$f$ lens configuration and a half-wave plate within a Sagnac common path interferometer \cite{sagnac1913comptes,post1967revmodphy,hariharan2010,naik2009optexp}. The incoming field is split by the beam splitter into two paths (indicated by two different shades of red), each traversing through the $4f$ lens configuration. Thus, there are two different alternative ways by which the field entering the interferometer reaches the detector plane of the EMCCD camera. Using two lenses of different focal lengths $f_1$ and $f_2$ in a $4f$ configuration, we magnify the fields in the two alternatives within the interferometer by different proportions, $m=f_1/f_2$ for field in path 1 (light red) and $f_2/f_1=1/m$ for the field in path 2 (dark red). This way, at the EMCCD camera plane, which is at the focal plane of both the lenses, we get a superposition of fields with different radial magnifications. The translation stage is used for alignment purposes and to maintain the $4f$ configuration of the lenses.  We insert a half-wave plate to rotate the linear polarization directions of the counter-circulating beams and record the interferograms corresponding to different relative polarization directions. The field in path 1 gets transmitted at the beam splitter twice, whereas the field in path 2 undergoes two reflections at the beam splitter. Therefore, for an input electric field $E_{\text {in}}(r, \theta)$ polarized along the horizontal direction, the electric field at the output port of the interferometer is given by:
\begin{align}
		\boldsymbol{E}_{\rm {out}}(r, \theta)=|r|^2e^{2i\phi_r} E_{\text {in}}\left(\frac{r}{m}, \theta\right) \hat{k}_1+ |t|^2e^{2i\phi_t} E_{\text {in}}\left({r}{m}, \theta\right) \hat{k}_2, \label{eqn:E_out}
\end{align}
Here, $r$ and $t$ are the reflection and transmission coefficients of the beam splitter, with $\phi_r$ and $\phi_t$ being the corresponding phases, respectively. The unit vectors $\hat{k}_1$ and $\hat{k}_2$ represent the polarization of the two interfering fields at the output port of the beam splitter. The half-wave plate (HWP) is oriented at angle $\phi$ with respect to the horizontal axis. As a result, it rotates the polarisation of the fields in path 1 and path 2 by $2\phi$ and $-2\phi$, respectively. Thus we get $\hat{k}_1=\cos 2 \phi \hat{x}+\sin 2 \phi \hat{y}$ and $\hat{k}_2=\cos 2 \phi \hat{x}-\sin 2 \phi \hat{y}$. Substituting for $\hat{k}_1$ and $\hat{k}_2$ in Eqn.~\ref{eqn:E_out}, we get
\begin{multline}
\boldsymbol{E}_{\text{out}}\left(r, \theta \right)=  \cos 2\phi\left[|r|^2e^{2i\phi_r} E_{\text {in}}\left(\frac{r}{m}, \theta\right)+ |t|^2e^{2i\phi_t} E_{\text {in}}\left({r}{m}, \theta\right)\right]\hat{x}\\
			 +\sin 2\phi\left[|r|^2e^{2i\phi_r} E_{\text {in}}\left(\frac{r}{m}, \theta\right)- |t|^2e^{2i\phi_t} E_{\text {in}}\left({r}{m}, \theta\right)\right]\hat{y}
\end{multline}
The intensity at the output port $I_{\text {out }}^\phi(r, \theta)=\left\langle \boldsymbol{E}_{\text {out }}^*(r, \theta) \boldsymbol{E}_{\text {out }}(r, \theta)\right\rangle_e$ can be shown to be:
\begin{align}
I_{\text {out }}^\phi(r, \theta)= |r|^4I_{\text {in}}\left(\frac{r}{m}, \theta\right)+|t|^4I_{\text {in}}\left(rm, \theta\right) - 2 |r|^2|t|^2\operatorname{Re}\left[ W\left(\frac{r}{m}, \theta, rm, \theta\right)\right] \cos (4\phi),
\end{align}
where $W(\frac{r}{m},\theta,rm,\theta)=\langle E_{\text {in}}^*(\frac{r}{m},\theta)E_{\text {in}}(rm,\theta)\rangle_e$, $I_{\text {in}}\left(\frac{r}{m}, \theta\right)=\langle E_{\text {in}}^*(\frac{r}{m},\theta)E_{\text {in}}(\frac{r}{m},\theta)\rangle_e$, etc.  $\operatorname{Re}\left[W\left(\frac{r}{m}, \theta,{r}{m}, \theta\right)\right]$  represents the real part of $W\left(\frac{r}{m},\theta, {r}{m}, \theta\right)$ and $\langle\cdots\rangle_e$ represents ensemble averaging. In deriving the above expression, we have used the fact that phases $\phi_r$ and $\phi_t$ in a beam splitter are related as $\phi_r-\phi_t=\dfrac{\pi}{2}$ \cite{degiorgio1980amjphy, henault2015quantum}. Thus, the real part of the cross spectral density function can be extracted by recording two interferograms with HWP kept at $\phi=0$ and $\pi/4$, as:
\begin{equation}
{\rm Re}\left[W\left(\frac{r}{m}, \theta, {r}{m}, \theta\right)\right] \propto\left[I_{\text {out }}^{\phi=\pi/4}(r, \theta)-I_{\text {out }}^{\phi=0}(r, \theta)\right].
\end{equation}
Here, $I_{\text {out }}^{\phi=0}(r, \theta)$ and $I_{\text {out }}^{\phi=\pi/4}(r, \theta)$ are the interferogram intensity at $\phi=0$ and $\phi=\pi/4$, respectively.	In this work, we only consider the cross spectral density functions that are real with no imaginary parts, such that ${\rm Re}[W\left(\frac{r}{m}, \theta, {r}{m}, \theta\right)]=W\left(\frac{r}{m}, \theta, {r}{m}, \theta\right)$. Thus, the radial cross-spectral density function $W_R\left(\frac{r}{m}, {r}{m}\right)$ of such fields can be obtained by performing the azimuthal averaging on $W\left(\frac{r}{m},\theta, {r}{m}, \theta\right)$, and it can be expressed in terms of the recorded interferograms as: 
\begin{equation}
W_R\left(r/m, {r}{m}\right) \propto \frac{1}{2\pi}\int_{-\pi}^{\pi}\left[I_{\text {out }}^{\phi=\pi/4}(r, \theta)-I_{\text {out }}^{\phi=0}(r, \theta)\right]d\theta.
\end{equation}
In the next two sections, we report the experimental measurements of radial cross-spectral density function for various different fields.
	
	\section{Experimental setup}
	\begin{figure*}[t!]
		\includegraphics[width =\columnwidth]{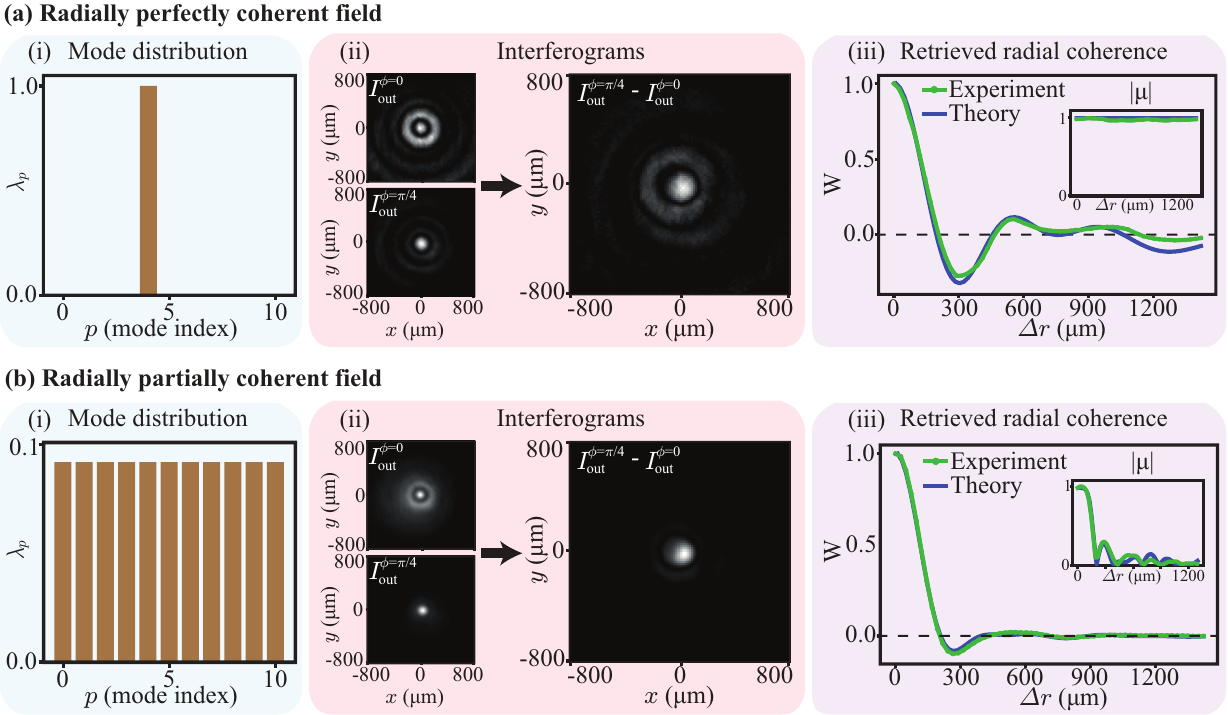}
		\caption{\label{fig:lg_both} (a) Radially perfectly coherent field $LG^{l=0}_{p=4}(r)$: (i) Mode distribution. (ii) Measured interferograms at $\phi=0$ and $\pi/4$ and their difference. (iii) Retrieved radial cross-spectral density function $W_R(r/2,2r)$ and degree of coherence function $|\mu_R(r/2,2r)|$ (inset). (b) Incoherent mixture of $LG^{l=0}_p(r)$ modes: (i) The mode distribution. (ii) Measured interferograms at $\phi=0$ and $\pi/4$ and their difference. (iii) Retrieved radial cross-spectral density function $|W_R(r/2,2r)|$ and degree of coherence function $|\mu_R(r/2,2r)|$ (inset).}
	\end{figure*}
	Figure \ref{fig:setup} shows the experimental setup for generating fields with tunable degree of coherence, and subsequently measuring it through the modified Sagnac interferometer. A 5mW He-Ne laser emits a Gaussian beam of central wavelength 632 nm, which illuminates a Holoeye Pluto spatial light modulator (SLM). We display a complex amplitude hologram on the SLM to generate a radial mode, $\mathrm{LG}_p^{l=0}(r)$ or $HG_{p}(r)$. Different modes are displayed for different time duration proportional to the coefficient $\lambda_p$ (see Eqn.~\ref{eqn:lg} or Eqn.~\ref{eqn:hg}). We set the exposure time of the EMCCD camera equal to the total duration time of all the holograms, ensuring that the camera detects the incoherent mixture of radial modes \cite{bhattacharjee2019jop}. 
	
	Each generated mode is directed through the modified Sagnac interferometer as shown in Fig.~\ref{fig:setup} for measuring the radial cross-spectral density function. The 4$f$ lens configuration within the interferometer consists of lenses with focal lengths $f_1=300$ mm and $f_2=150$ mm, resulting in a magnification factor of $m=2$. Consequently, the incoming field is magnified 2 times in path 1, while it is demagnified by the same proportion in path 2. The translation stage is used for alignment purposes and to maintain the $4f$ configuration of the lenses. Finally, we record the interferograms for HWP angles $\phi=0$ and $\pi/4$ using the EMCCD camera.

	\section{Results}
	\subsection{Incoherent mixture of Laguerre-Gaussian (LG) modes}
	
	As a proof-of-principle, we first apply our scheme to characterise a Laguerre-Gaussian (LG) $LG^{l=0}_p(r)$ mode, for $p=4$, which is a radially perfectly coherent field. Consequently, the mode distribution $\lambda_p$ (see Eqn. \ref{eqn:lg}) in Fig.~\ref{fig:lg_both}(a)(i) shows a single mode. Fig.~\ref{fig:lg_both}(a)(ii) shows the measured interferograms for HWP angles $\phi=0$ and $\phi=\pi/4$, and their intensity difference. We extract the radial cross-spectral density function $W_R(r/2,2r)$ from the measured interferograms by following the steps described in Sec. \ref{sec:scheme}. Fig.~\ref{fig:lg_both}(a)(iii) shows the retrieved $W_R(r/2,2r)$ as a function of separation $\Delta r\equiv 2r-r/2=3r/2$ alongside the theoretical profile. Subsequently, we measure the radial intensity distribution $I(r)$ of the field using the EMCCD camera. By combining $I(r)$ with the measured $W_R(r/2,2r)$ for $m=2$ in Eqn.~\ref{mu_eq}, we retrieve the degree of coherence profile $|\mu_R(r/2,2r)|$ as a function of $\Delta r$ shown as an inset in Fig. \ref{fig:lg_both}(a)(iii). The degree of coherence profile $|\mu_R(r/2,2r)|$ remains close to 1 for any radial separation $\Delta r$, confirming the perfect coherence between different radial locations.
	
	Next, we generate a radially partially coherent field by incoherently mixing LG modes $LG^{l=0}_p(r)$ with different $p$ indices ($p=0$ to $p=10$). The weight distribution $\lambda_p=1/11$ (see Eqn.~\ref{eqn:lg}) for this incoherent mixture is shown in Fig.~\ref{fig:lg_both}(b)(i). The measured interferograms at $\phi=0$ and $\phi=\pi/4$, along with the corresponding difference, are shown in Fig.~\ref{fig:lg_both}(b)(ii). The radial cross-spectral density function $W_R(r/2,2r)$ and degree of coherence function $|\mu_R(r/2,2r)|$ (inset) are shown in Fig.~\ref{fig:lg_both}(b)(iii) alongside their theoretical predictions. In contrast to the radially perfectly coherent field, the degree of coherence profile $|\mu_R(r/2, 2r)|$ decays with increasing radial separation $\Delta r$, demonstrating how the coherence between two radial positions deteriorates as their separation increases.

	\subsection{Incoherent mixture of radial Hermite-Guassian (HG) modes}
	\begin{figure*}[t]
		\includegraphics[width =\columnwidth]{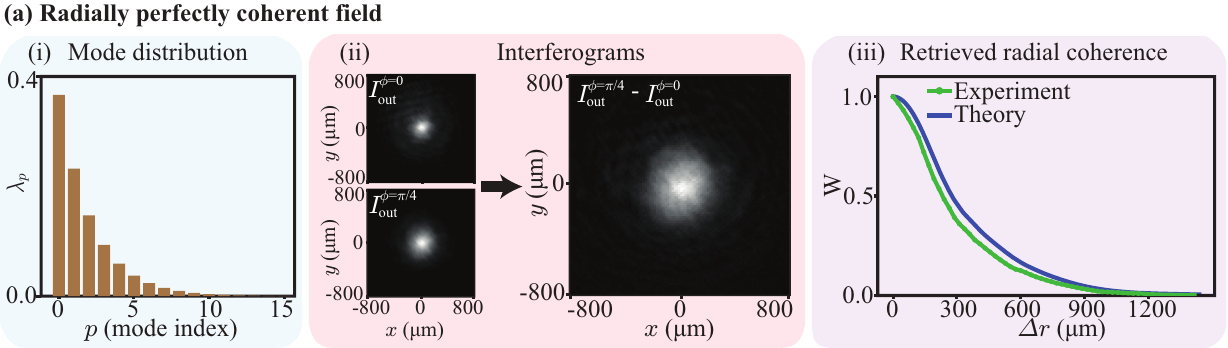}
		\caption{\label{fig:hg_sum} Radial Gaussian Schell Model. (i) Mode distribution. (ii) Measured interferograms at $\phi=0$ and $\pi/4$ and their difference. (iii) Retrieved radial cross-spectral density function $W_R(r/2,2r)$.}
	\end{figure*}
	We generate a set of radial Hermite-Gaussian (HG) fields $HG_p(r)$ ($p=0$ to $p=15$) and incoherently mix them with weights $\lambda_p$ (see Eqn. \ref{eqn:hg}) corresponding to $Q=0.5$ (as shown in Fig. 3(i)) for producing a radial Gaussian Schell Model (GSM) field \cite{bhattacharjee2023josaa}.  
	
	Figure \ref{fig:hg_sum}(ii) shows the measured interferograms for $\phi=0$ and $\phi=\pi/4$, and their difference. From these interferograms, we retrieve radial cross-spectral density function $W_R(r/2,2r)$, which is plotted as a function of $\Delta r$ alongside the theoretical prediction in Fig.~\ref{fig:hg_sum}(iii). The retrieved $W_R(r/2,2r)$ shows the expected Gaussian profile, a characteristic of the radial GSM field.

Our measured results show very good agreement with theoretical predictions, which highlights the capability of this scheme for characterising a field with arbitrary degree of radial coherence. Minor mismatch between the experimental and theoretical results can be attributed to imperfections in generating radial modes and inaccuracies in the weight distribution. Additionally, slight deviation in the distances within the $4f$ lens configuration can introduce a relative spherical wavefront difference between the two interferometric paths. This can be overcome by installing the $4f$ lens configuration on a precisely calibrated translation stage.

	\section{Conclusions and Discussions}
	
We have demonstrated an efficient experimental technique for measuring radial coherence. We have reported measurements of radial coherence of two different types of radially partially coherent fields. The first type is the class of fields that has the coherent mode representation in the LG basis, and the second type is the class of fields having coherent mode representation in the radial HG modes. Measurement of radial coherence should yield radial mode spectrum in an analogous manner as the OAM mode spectrum is obtained from the angular coherence measurements. The radial degree of freedom in combination with the angular degree of freedom can be leveraged for providing very high-dimensional single-photon states for quantum information applications.
	
	\begin{backmatter}
		\vspace{-2 pt}
		\bmsection{Funding}
Science and Engineering Research Board (STR/2021/000035, VJR/2021/000012); Science and Engineering Research Board (CRG/2022/003070); Department of Science and Technology, Ministry of Science
and Technology, India (DST/ICPS/QuST/Theme-1/2019).		
				\vspace{-2 pt}
		\bmsection{Acknowledgment}
		R.P. and N.S. thank the Prime Minister's Research Fellowship, Ministry of Education, Government of India for financial support. M.A.A. acknowledges the hospitality of IIT Kanpur’s Physics Department and financial support during his visit under the VAJRA Faculty Scheme of India’s Science and Engineering Research Board. B. P. acknowledges financial support of PNRR MUR project PE0000023-NQSTI and PRIN MUR project CUP E53D23002520006.
		\vspace{-2 pt}
		\bmsection{Disclosures}
		The authors declare no conflicts of interest.
		\vspace{-2 pt}
		\bmsection{Data Availability Statement}
		The data presented in the paper can be obtained from the authors upon reasonable request.
	\end{backmatter}
\vspace{-5 pt}
\bibliography{ref_oe_new}

\end{document}